\documentclass{PoS}

\title{Non abelian Bianchi identities, monopoles and gauge invariance}

\ShortTitle{NABI, monopoles and gauge invariance}

\author{\speaker{Claudio Bonati}\\
        Dipartimento di Fisica, Universit\`a di Pisa and INFN, Largo Pontecorvo 3, I-56127 Pisa, Italy.\\
        E-mail: \email{claudio.bonati@pi.infn.it}}

\author{Massimo D'Elia\\
        Dipartimento di Fisica, Universit\`a di Genova and INFN, Via Dodecaneso 33, 16146 Genova, Italy.\\
        E-mail: \email{massimo.delia@ge.infn.it}}

\author{Adriano Di Giacomo\\
        Dipartimento di Fisica, Universit\`a di Pisa and INFN, Largo Pontecorvo 3, I-56127 Pisa, Italy.\\
        E-mail: \email{digiaco@df.unipi.it}}

\author{Luca Lepori\\
        Scuola Internazionale Superiore di Studi Avanzati and INFN,Via Beirut 2-4, 34151 Trieste, Italy.\\
        E-mail: \email{lepori@sissa.it}}

\author{Fabrizio Pucci\\
        Dipartimento di Fisica, Universit\`a di Firenze and INFN, Via Sansone 1, 50019 Sesto Fiorentino, Firenze, Italy.
        \thanks{Present address: Fakult\"{a}t f\"{u}r Physik, Universit\"{a}t Bielefeld,  D-33615 Bielefeld, Germanyy.} \\
        E-mail: \email{pucci@physik.uni-bielefeld.de}}

\abstract{A direct connection is proved between the Non-Abelian Bianchi Identities and the Abelian Bianchi identities for the 't Hooft tensor in a generic gauge; the existence of a magnetic current is related to the violation of NABI's. Using this relation it is shown that not all gauges are equivalent to detect monopoles on the lattice, that e.g. the Maximal Abelian Gauge is a legitimate choice while the Landau gauge is not. Nevertheless monopole condensation is found to be a gauge invariant property.}

\FullConference{The XXVIII International Symposium on Lattice Filed Theory\\
		 June 14-19,2010\\
		 Villasimius, Sardinia Italy}

\def\Tr{\mathrm{Tr}} 
\def\[{\lbrack}
\def\]{\rbrack}
\def\eg{\emph{e.g.} }

\def\vev{\emph{vev} }
\newcommand{\eqref}[1]{Eq.~(\ref{#1})}

\begin{document}

\section{Introduction}
Monopoles have been proposed to be the excitations of QCD responsible for the  confinement of color, which, 
in this picture, is interpreted as a consequence of the dual superconductivity of the vacuum induced by the 
monopole condensation \cite{mandelstam, thooft77}. 

Because of this proposal, a big activity has developed during the years in the lattice community to detect 
and study monopoles in numerically generated QCD configurations. These studies can be divided into two main groups
with different working strategies:
\begin{itemize}
\item the study of the (dual) symmetry of the vacuum state
\item the direct observation of monopoles in lattice configurations
\end{itemize} 

In the studies of the first group an order parameter for dual superconductivity is introduced: this is defined as the 
vacuum expectation value (\vev$\!\!\!$) of a magnetically charged operator $\mu$. $\langle\mu\rangle$ is zero if the magnetic 
charge is superselected and nonzero in the dual superconducting phase (see \eg \cite{digz, ddpp, vpc, dlmp, ccos, 3} and references therein). 
The works of the second group try instead to extract a monopole effective action or to check monopole dominance 
by direct detection of monopoles in lattice configurations (see \eg \cite{suz, pol, scol} and references therein). 

In this paper we shall analyze some general properties of monopoles in non abelian gauge theories, whose applications 
are more direct in the studies of the second group but that are also interesting from the point of view of the works of the
first one.

The standard method to detect monopoles in lattice configurations was first introduced in Ref.~\cite{dgt} for the 
$U(1)$ gauge theory: a Dirac string is identified by any excess over $2\pi$ of the abelian phase of a plaquette and a 
monopole is located inside an elementary cube whenever a net number of Dirac strings crosses its boundary plaquettes. This
procedure is well defined since in $U(1)$ gauge theory the abelian phase of a plaquette is gauge invariant. For non abelian
gauge theories this recipe must be modified: one has first to fix a gauge, and then to apply the above prescription to the
components of the non abelian field directed along an abelian subgroup of the original gauge group (abelian projection).

While for the 't Hooft-Polyakov monopole of Ref.~\cite{thooft74, polyakov} the magnetic $U(1)$ group is naturally identified with the 
little group of the Higgs \vev in the unitary representation, in gauge theories without Higgs field there is no obvious preferred 
gauge direction. Indeed it was proposed in Ref.~\cite{thooft81} that any operator in the adjoint representation of the group could be used 
as an effective Higgs field to identify the magnetic $U(1)$, physics being for some reason independent of that choice. It 
was however observed in numerical simulations that the number and the position of the monopoles detected in a given 
configuration was strongly dependent on the abelian projection.

For the interpretation of the color confinement as dual superconductivity it is crucial to be able to define monopoles 
in a gauge invariant way. We will show that this is possible by using the Non Abelian Bianchi Identities (NABI) and that
for any magnetically charged gauge field configuration a preferred direction in color space exists, which identifies the
correct magnetic $U(1)$ subgroup. As a consequence monopoles are gauge invariant objects but their detection by the recipe
of Ref.~\cite{dgt} it is not; this techniques works well for a class of gauges which include the maximal abelian gauge but
it fails for the Landau gauge, where no monopoles are detected. Dual superconductivity, defined as the breaking of the
magnetic $U(1)$, is also gauge invariant.

\section{Abelian and Non-Abelian Bianchi Identities}

Abelian Bianchi identities are $\partial_{\mu}F_{\mu\nu}^*=0$, where $F_{\mu\nu}^*=\frac{1}{2}\epsilon_{\mu\nu\rho\sigma}
F_{\rho\sigma}$ is the dual of the abelian field strength, and their violation naturally defines the magnetic current 
\begin{equation}\label{abi}
j_{\nu}=\partial_{\mu}F_{\mu\nu}^*
\end{equation}
which is conserved because of the antisymmetry of the $F_{\mu\nu}^*$ tensor: $\partial_{\nu}j_{\nu}=0$. The Non Abelian
Bianchi Identities are the covariant generalization of the abelian ones for non abelian gauge theories and  
their violation defines the current 
\begin{equation}\label{nabi}
J_{\nu}=D_{\mu}G_{\mu\nu}^*
\end{equation}
which is not difficult to show to be covariantly conserved: $D_{\nu}J_{\nu}=0$.

In non abelian gauge theories the abelian field strength $F_{\mu\nu}$ is defined by means of the 't Hooft tensor, which 
reduces to the abelian field strength of the residual $U(1)$ in the unitary gauge; its introduction goes back to 
Ref.~\cite{thooft74} for the group $SU(2)$, while the generalization for an arbitrary compact gauge group was developed in 
Ref.~\cite{dlp}.

Because of the Coleman-Mandula theorem, the four components of the current $J_{\nu}$ defined in \eqref{nabi} commute with
each other; to expose the gauge invariant content of this equation is thus possible to simultaneously diagonalize 
all its component. A convenient basis for the diagonal matrices is that of the fundamental weights $\phi_0^a$, 
$a=1,\ldots,r$, where $r$ is the rank of the gauge group. A fundamental weight $\phi_0^a$ is associated to each of 
the simple roots $\vec{\alpha}^a$ of the group algebra; its commutations rules with the elements of the Cartan base are
$\[\phi_0^a, H_i\]=0$ and $\[\phi_0^a, E_{\pm\vec{\alpha}}\]=\pm(\vec{c}^{~a}\cdot \vec{\alpha})E_{\pm\vec{\alpha}}$ and
for the simple roots $\vec{c}^{~a}\cdot\vec{\alpha}^b=\delta^{ab}$. Let $\phi_I^a$ be the adjoint representation operator
equal to $\phi_0^a$ in the gauge that diagonalize $J_{\nu}$; then the gauge invariant content of \eqref{nabi}
is
\begin{equation}
\Tr\left(\phi_I^a D_{\mu}G_{\mu\nu}^*\right)=\Tr\left(\phi_I^a J_{\nu}\right)
\end{equation}

Let us now denote by $\phi^a$ the matrix which is equal to $V(x)\phi_0^aV(x)^{\dag}$ in the gauge in which $J_{\nu}$ 
is diagonal, $V(x)$ being a generic gauge transformation. It was proved in Ref.~\cite{bdlp} that, for a generic compact 
gauge group, the abelian magnetic current $j_{\nu}^{\,a}$ in the abelian projection defined by the operator $\phi^a$ 
satisfies
\begin{equation}\label{th}
\partial_{\mu}F_{\mu\nu}^{\, a\,*}= j_{\nu}^{\, a}=\Tr\left(\phi^a J_{\nu}\right)
\end{equation}
where $F_{\mu\nu}^{\, a\,*}$ is the 't Hooft tensor in the given abelian projection.
This equation shows that the existence of a magnetic current is related to the violations of the NABI's and that its
value is just the abelian projection of that violation, thus revealing the hidden gauge invariance of the 
monopole's definition.

\section{The 't Hooft-Polyakov monopole}

In this section we will check the equation \eqref{th} for the soliton solution of Ref.~\cite{thooft74, polyakov} and we
will make some simple observations whose relevance will become clear in the next sections.

The solution of Ref.~\cite{thooft74, polyakov} can be written in the form
\begin{equation}\label{tp_hedg}
\phi^a(\vec{r})=H(r)\frac{r^a}{r} \qquad A_0=0\qquad A_{i}^a=\epsilon_{iak}\frac{r^k}{gr^2}[1-K(r)]
\end{equation}
where $H$, $K$ are two functions whose specific form depends on the details of the Higgs potential. Their universal
features are their asymptotics: $\lim_{r\to 0}H(r)=0$ and $1-K(x)\propto x^2$ for small $x$ in order to ensure regularity;
$\lim_{r\to\infty}H(r)=const$ and $\lim_{x\to\infty}K(x)=0$ to have a topologically stable solution of finite total energy.

The gauge in which the solution \eqref{tp_hedg} is written is usually called the ``hedgehog'' one and it is trivial to 
verify that it satisfies the Landau gauge condition $\partial_{\mu}A_{\mu}=0$. The asymptotic abelian magnetic field at large 
distance in this gauge can be shown to be given by the expression
\begin{equation}
b^i(\vec{r})\approx \frac{r^i}{gr^2}\frac{z}{r}
\end{equation}
and thus the magnetic charge $Q_m$, calculated by the flux at infinity, is zero.

Let us now consider the unitary gauge, that is the gauge in which the Higgs field is rotated into a fixed direction
in color space. The explicit form of the gauge field can be computed starting from \eqref{tp_hedg} 
(see \eg Ref.~\cite{shnir}) and it is simple to verify that the solution in the unitary gauge satisfies the equation
\begin{equation}\label{mag}
\partial_{\mu} A^{\pm}_{\mu} + ig \left[A^3_{\mu}, A^{\pm}_{\mu}\right] = 0 
\end{equation}
which is the continuum form of the maximal abelian gauge (MAG) introduced in Ref.~\cite{sch}. In this gauge it is easy
to show that the temporal component (the only non-vanishing one) of the non abelian current \eqref{nabi} is given by
\begin{equation}
J_0=D_iB_i=\frac{2\pi}{g}\delta^3(\vec{r})\sigma_3
\end{equation}
which is diagonal in color space: the unitary gauge coincides with the abelian projection 
indicated by $\phi_I^a$ in the previous section. For $SU(2)$ there is only one fundamental weight, 
$\phi_0=\frac{1}{2}\sigma_3$, associated to the $+1$ simple root. By direct calculation it can also be
shown that the temporal component of the abelian magnetic current is equal to
\begin{equation}
j_0=\vec{\nabla}\cdot\vec{b}=\frac{2\pi}{g}\delta^3(\vec{r})
\end{equation}
consistently with the theorem of the previous section, \eqref{th}. In the unitary gauge the magnetic charge is thus 
$Q_m=\frac{1}{2g}$; since the elementary electric charge is given by $Q_e=\frac{g}{2}$, $Q_m$ is equal to two Dirac units.

To summarize we have shown that monopoles are related to NABI's violations and they are thus gauge invariant objects; 
however, because of the presence of a preferred direction in color space (that of the NABI's violation), the magnetic 
charge computed by using the flux at infinity depends on the gauge choice. In particular we have just shown that, for 
the 't Hooft-Polyakov soliton, the magnetic charge in the Landau gauge is zero, while in the maximal abelian gauge 
it is equal to two Dirac units.

\section{General monopole configuration}

The conclusions of the previous section can be generalized to an arbitrary configuration by using the following theorem
of Ref.~\cite{coleman}: \emph{the magnetic monopole term in the multipole expansion of a generic static field configuration
is abelian (i.e. satisfies abelian equations of motion) and can be gauged along a fixed direction in color space}.

Up to a global gauge transformation we can suppose the asymptotic non abelian magnetic field of the configuration $A_{\mu}$ to be directed 
along the $3-$axis in color space; at large $r$ it is then fixed by the total magnetic charge $m$ (in Dirac units) to be
\begin{equation}
\vec{B}=\frac{m}{2}\frac{\vec{r}}{2g r^3}\sigma_3
\end{equation}
By adding and subtracting to the gauge field a term $\tilde{A}_{\mu}$ given by the 't Hooft-Polyakov solution in the unitary 
gauge, multiplied by $\frac{m}{2}$, we can see the field $A_{\mu}$ as a superposition of a 't Hooft-Poliakov-like term $\tilde{A}_{\mu}$
and a topologically trivial term $\tilde{A}_{\mu}^{(0)}=A_{\mu}-\tilde{A}_{\mu}$. Since the 't Hooft tensor is linear in 
the gauge field, the asymptotic abelian magnetic field of $\tilde{A}_{\mu}^{(0)}$ is zero, and the results of the previous section 
can extended to a general static configuration: in the MAG, or in any gauge which differs from it by an asymptotically 
trivial gauge transformation, the flux of the abelian magnetic field at large distance gives the correct magnetic charge. In other gauges 
the flux at infinity will be generally smaller than the MAG one and for the Landau gauge it will be vanishing. 

For a non static configuration, the same argument can be applied to the superposition of the given configuration and its
time-reversed one, in order to isolate the asymptotic magnetic field from the electric field.

\section{Monopole condensation}

From equation \eqref{th} it follows that in $SU(2)$ gauge theory the magnetic charge in the MAG is given by
\begin{equation}
Q_I=\int \mathrm{d}^3x\, \Tr(\phi_I J_0(x)) \label{charge}
\end{equation}
A magnetically charged operator $O(x)$ satisfies the commutation rule
\begin{equation}
\[Q_I, O(x)\]=m\, O(x) \qquad m\ne 0
\end{equation}
and the magnetic $U(1)$ symmetry is broken if a charged operator exists for which $\langle O\rangle\ne 0$. We will show now that this 
property is gauge invariant: in a generic abelian projection $V$ the magnetic charge $Q_V$ has the form in \eqref{charge} with 
$\phi_I\to V(x)\phi_I V(x)^{\dag}$ and, since the integrand is gauge invariant, we can compute the trace in the gauge in which $J_0$ and $\phi_I$
are diagonal. Since $V(x)\phi_I V(x)^{\dag}$ is an element of the group algebra it can be expanded in the Cartan base:
\begin{equation}
V(x)\phi V(x)^{\dag}=C(x,V)\phi_I + \sum_{\vec{\alpha}}E_{\vec{\alpha}}D^{\vec{\alpha}}(x,V) \label{expansion}
\end{equation}
In the chosen gauge only the first term contributes and thus 
\begin{equation}
\[Q_I, O(x)\]=m\,C(x,V)\, O(x)
\end{equation}
Since $C(x,V)$ is generically non-vanishing, the operator $O$ will be charged also with respect to $Q_V$ and the corresponding $U(1)$ magnetic 
symmetry will be broken. For a general gauge group the only difference is that in \eqref{expansion} the term $C(x,V)\phi_I$ has to be replaced by
a sum over the fundamental weights $\phi_I^a$ ($a=1,\ldots,r$).

\section{Monopoles detection on the lattice} 

The detection of monopoles in lattice configurations is performed by measuring the abelian magnetic flux, in a given abelian projection, through the 
surface of elementary cubes \cite{dgt}. The previous analysis shows that the magnetic flux depends on the chosen abelian projection: the ``correct'' 
magnetic charge is measured in the MAG, while in other gauges it is expected to be smaller. 

A direct test of the previous picture is presented in Ref.~\cite{bdd}. The starting point is a configuration in the MAG; in 
this gauge it is known that the flux on the boundary of an elementary cube gives a reliable estimate of the flux at infinity (see Ref.~\cite{ddmo}), 
so we can safely apply the recipe of Ref.~\cite{dgt} to locate monopoles. If we now assume that the monopole is at the center of a cube and the 
Dirac string along the $z$ direction, we can perform the gauge transformation 
\begin{equation}
U(a)=\exp\left(-i\phi\frac{\sigma_3}{2}\right)\exp\left(-ia\theta\frac{\sigma_2}{2}\right)\exp\left(i\phi\frac{\sigma_3}{2}\right) \label{trans}
\end{equation}
where $\theta, \phi$ are the polar angles and $0\le a\le 1$ is a free parameter. For $a=0$ one stays in the MAG, for $a=1$ the transformation in 
\eqref{trans} is the unitary matrix bringing from the MAG to the Landau gauge (see \eg \cite{shnir}). In the gauge defined by the parameter $a$, 
the magnetic charge, defined by the abelian magnetic flux at infinity, is given by
\begin{equation}
\frac{Q(a)}{Q(0)}=\frac{1+\cos(a\pi)}{2} \label{q_ratio}
\end{equation} 
The comparison between \eqref{q_ratio} and the lattice measurements was performed in Ref.~\cite{bdd}. Although in its derivation we completely neglected
all sort of discretization errors, \eqref{q_ratio} qualitatively well describes the observed behaviour of $Q(a)$.

\section{Conclusions}

We have shown that monopoles are gauge invariant objects and that each configuration with non-zero magnetic field 
selects a natural direction in color space, induced by the violations of the non-abelian Bianchi identities. 
This direction 
is correctly identified by the maximal abelian gauge, which is the gauge to be used to 
detect monopoles in lattice configuration by the DeGrand-Toussaint recipe. While monopole condensation is a gauge invariant fact, monopole detection 
is strongly affected by the gauge choice in a well understood way.

\end{document}